%Paper: hep-th/9207083
%From: "Paul A. Griffin ph. (904)392-5712" <pgriffin@ufhepa.phys.ufl.edu>
%Date: Fri, 24 Jul 1992 15:26:28 EDT

\input harvmac.tex
%%%%% 2 FIGURES AND EPSF.TEX AVAILABLE VIA ANONYMOUS FTP TO
%%%%% uful07.phys.ufl.edu IN DIRECTORY 	het/UFIFT-HEP-92-19
%
\def\havefigures{n}
% if you have the figures, epsf.tex, and the dvips dvi --> postscript
% converter, then uncomment the following line:
%\def\havefigures{y}

%%%%% figure macro ( \topfigure#1#2#3 )
% #1 = figure.eps file, #2 = figure number, #3 = figure caption
% place \topfigure command in separate paragraph

\def\figures{y}
\ifx\havefigures\figures
	\input epsf.tex
	\def\topfigure#1#2#3{\topinsert {\centerline{\epsffile{#1}}}
	\smallskip\tenpoint\baselineskip12pt \noindent
	{\bf Fig.~{#2}.} {\rm \ \  {#3}} \endinsert}
\else
	\def\topfigure#1#2#3{}
\fi

%%%%% title page macros

\def\preprint#1{\nopagenumbers\abstractfont\pageno=0
\hsize=\hstitle\rightline{#1}}

\def\date#1{\rightline{#1}}

\def\title#1#2{
   \vskip 1in
   \centerline{\titlefont #1}
   \vskip 0.2in
   \centerline{\titlefont #2}
   \vskip 0.5in plus 0.1in}

\def\author#1#2#3{\smallskip\centerline{{\bf #1}\footnote{#2}{#3}}\smallskip}
\def\address#1{\centerline{#1}}

\def\abstract{\centerline{\bf Abstract}\smallskip}
\def\finishtitlepage{\Date{\ }}

%%%%%  \draft that doesn't change line spacings

\def\monthintext{\ifcase\month\or January\or February\or
   March\or April\or May\or June\or July\or August\or
   September\or October\or November\or December\fi}

%%%%% footnote

\let\ulabelfoot=\footnote

%%%%% \subsection ( \subsec )

\def\undersection#1{\par
   \ifnum\the\lastpenalty=30000\else \penalty-100\medskip \fi
   \noindent\undertext{#1}\enspace \vadjust{\penalty5000}}
\def\undertext#1{\vtop{\hbox{#1}\kern 1pt \hrule}}
\def\subsection#1{\undersection{#1} \medskip}

%%%%% acknowledgements

\def\ack{\ifnum\the\lastpenalty=30000\else \penalty-100\smallskip \fi
   \noindent\undertext{Acknowledgements:}\enspace \vadjust{\penalty5000}}

%%%%%  personal definitions

\def\en{\eqalign}
\def\del{\partial}

%%%%%  end of harvmac_post.tex  %%%%%%%%%%%%%%%%%%%%%%%%%%%%%%%%%%%%%%%

%xsb.hea -- definitions and references
%% Definitions
\def\en{\eqalign}

\def\xperp{ {{\vec x}_{\perp}} }
\def\yperp{ {{\vec y}_{\perp}} }

\def\bfalpha{ {\vec \alpha}}
\def\svec{ {\vec s} }
\def\yhat{ {\vec y} }
\def\xhat{ {\vec x} }

\def\cL{ {\cal L} }
\def\cO{ {\cal O} }
\def\cG{ {\cal G} }
\def\cH{ {\cal H} }
\def\cN{ {\cal N} }
\def\cM{ {\cal M} }

\def\tphi{ {\tilde \phi} }
\def\psibar{ {\bar \psi} }
\def\state#1{ {| #1\rangle} }
\def\lstate#1{ {\langle #1 |} }

\def\rvac{ {| 0 \rangle} }
\def\lvac{ {\langle 0 |} }
\def\btl{\beta_{\rm tl } }
%
%% References
% section 1
\nref\suss{L.~Susskind, Phys.~Rev.~D16 (1977)3031.}
\nref\wilson{K.~Wilson, Phys.~Rev.~D10 (1974)2445.}
\nref\ks{J.~Kogut and L.~Susskind, Phys.~Rev.~D11 (1975)395.}
\nref\bp{W.~Bardeen and R.~Pearson, Phys.~Rev.~D14 (1976)547.}
\nref\bpr{W.~Bardeen, R.~Pearson, and  E.~Rabinovici,
Phys.~Rev.~D21 (1980)1037.}
\nref\miransky{V.A.~Miransky, Nuovo Cimento 90A (1985)149.}
\nref\llb{C.~Leung, S.~Love, and W.~Bardeen, Nucl.~Phys.~B273 (1986)649.}
\nref\seiler{M.~Salmhofer and E. Seiler, Comm.~Math.~Phys.~139 (1991)395.}
\nref\semenoff{G.W.~Semenoff, UBC preprint 92-0102, 1992.}
\nref\quenched{J.~Bartholomew et.~al., Nucl.~Phys.~B230~(1984)222.}
\nref\kdk{J.~Kogut, E.~Dagotto, and A.~Kocic, Nucl.~Phys.~B317
 (1989)253,271.}
\nref\dgroup{M.~G\"ocker, et.~al., Nucl.~Phys.~B334 (1990)527.}
\nref\ddgroup{M.~G\"ocker, et.~al., Phys.~Lett.~251B (1990)567.}
%section 3
%\nref\griffina{P.~Griffin, Mod.~Phys.~Lett.~A, March(1992).}
\nref\griffinb{P.~Griffin, Nucl.~Phys.~B372~(1992)270.}
\nref\smit{N.~Kawamoto and J.~Smit, Nucl.~Phys.~B192~(1981)100.}
\nref\kmnp{H.~Kluberg-Stern, A.~Morel, O.~Napoly, and B.~Petersson,\hfill\break
Nucl.~Phys.~B190~(1981)504.}
\nref\mpr{E.~Marinari, G.~Parisi, and C.~Rebbi,
Phys.~Rev.~Lett.~47~(1981)1795.}
\nref\hp{H.~Hamber and G.~Parisi, Phys.~Rev.~Lett.~47~(1981)1792.}
\nref\kmn{H.~Kluberg-Stern, A.~Morel, and O.~Napoly, Nucl.~Phys.~B220
(1983)447.}
%section 4
\nref\leibbrandt{G.~Leibbrandt, Rev.~Mod.~Phys.~47 (1975)849.}
\nref\colemanth{S.~Coleman, Ann.~Phys.~93 (1975)267.}
\nref\ksop{J.~Kogut and D.~Soper, Phys.~Rev.~D1 (1970)2901.}
\nref\bks{J.D.~Bjorken,J.~Kogut, and D.~Soper, Phys.~Rev.~D3 (1971)1382.}
\nref\cry{S.~Chang, R.~Root, and T.~Yan, Phys.~Rev.~D7 (1973)1133,1147.}
\nref\pinsky{D.~Mustaki, S.~Pinsky, J.~Shigemitsu,and K.~Wilson,
\hfill\break Phys.~Rev.~D43~(1991)3411.}
\nref\thorn{C.~Thorn, Phys.~Rev.~D20 (1979)1934.}
\nref\casher{A.~Casher, Phys.~Rev.~D14 (1976)452.}
\nref\bpauli{S.~Brodsky and C.~Pauli, Phys.~Rev.~D32 (1985)1993.}
\nref\mrobertson{G.~McCartor, D.~Robertson, Z.~Phys.~C53(1992)679.}
\nref\griffinsg{P.~Griffin, Univ. Florida preprint UFIFT-HEP-92-17, 1992.}
\nref\brodsky{S.~Brodsky and R.~Roskies, and R.  Suaya, Phys.~Rev.~D8
(1973)4574.}
\nref\colemansg{S.~Coleman, Phys.~Rev.~D11 (1975)2088.}
\nref\mandelstam{S.~Mandelstam, Phys.~Rev.~D11 (1975)3026.}
%section 5
\nref\samuel{S.~Samuel, Phys.~Rev.~D18 (1978)1916.}
\nref\mrg{P.~Minnhagen, A.~Rosengren, and G.~Grinstein,
Phys.~Rev.~B18(1978)1356.}
\nref\agg{D.~Amit, Y.~Goldschmidt, and G.~Grinstein, J.~Phys.~A13 (1980)585.}
%section 6
\nref\colthm{S.~Coleman, Comm.~Math.~Phys.~31 (1973)259.}
\nref\balaban{T.~Balaban, J.~Imbrie, A.~Jaffe, and
D.~Brydges,~Ann.~Phys.~158~(1984)281.}
\nref\lmattis{E.~Lieb and D.~Mattis, Jour.~Math.~Phys.~3 (1962)749.}
\nref\kls{T.~Kennedy, E. Lieb, and B. Sriram Shastry, Jour. Stat. Phys. 53
(1988)1019.}
\nref\gss{M.~Gross, E.~S\'anchez-Velasco, and E.~Siggia, Cornell preprint
88-0366 (1988).}

\preprint{UFIFT-92-19}
\date{July 1992}
\title	{Staggered fermions and chiral symmetry breaking}
	{in transverse lattice regulated
	QED\ulabelfoot{{\titlefont$^\dagger$}}
	{Supported in part by the U.S.  Department of Energy,
	under grant DE-FG05-86ER-40272} }
\author	{Paul A. {Grif}fin} {$^{\dagger\dagger}$}
	{Internet Address: p{grif}fin@ufhepa.phys.ufl.edu}
\address{Department of Physics, University of Florida}
\address{Gainesville, FL 32611}
\abstract
Staggered fermions are constructed for the transverse lattice
regularization scheme.  The weak perturbation theory of
transverse lattice non-compact QED is developed in light-cone gauge,
and we argue that for fixed lattice spacing this theory is ultraviolet finite,
order by order in perturbation theory.  However, by calculating the
anomalous scaling dimension of the link fields, we find that the interaction
Hamiltonian becomes non-renormalizable for $g^2(a) > 4\pi$, where $g(a)$ is the
bare (lattice) QED coupling constant.  We conjecture that this is the
critical point of the chiral symmetry breaking phase transition in QED.
Non-perturbative chiral symmetry breaking is then studied in the strong
coupling limit.  The discrete remnant of chiral
symmetry that remains on the lattice is spontaneously
broken, and the ground state to lowest order in the strong
coupling expansion corresponds to the classical ground
state of the two-dimensional spin one-half Heisenberg antiferromagnet.

\finishtitlepage

%% Introduction
\newsec{Introduction}

Staggered fermions\suss\ for lattice gauge theory \wilson\ks\ have the
desirable
property of preserving a discrete remnant of chiral symmetry, and are
therefore useful for studying the non-perturbative chiral symmetry breaking in
gauge theories.  Staggered fermions have been constructed
for the four dimensional Euclidean formulation of lattice gauge theory, and for
the Hamiltonian formulation of lattice gauge theory,
based on a three dimensional spatial lattice and one continuum time variable.
In sections 2 and 3, we construct staggered fermions for the
transverse lattice formalism of Bardeen  et.~al.\bp\bpr, which is based on
a two-dimensional spatial lattice and two continuum space-time coordinates.
Wilson fermions for the transverse lattice were constructed in ref.~\bp.

The transverse lattice construction is a minimalist's
non-perturbative regularization scheme for gauge fields\bp.
After choosing an axial gauge
and imposing the Gauss constraint, the degrees of freedom of the
gauge field are reduced to two spatial components, and these can be regulated
by mapping them to link fields on a two-dimensional
lattice.   The link fields are non-perturbative excitations of the
gauge fields,
and are scalars with respect to the two continuous space-time
coordinates perpendicular to the lattice, so their ultraviolet (UV) behavior is
softened.

The basic disadvantage of the transverse lattice construction is the breaking
of 3+1 dimensional Lorentz invariance down to 1+1 dimensional Lorentz
invariance plus discrete 2-D lattice translations and rotations.
This means, for example, that pure 3+1 dimensional gauge theory has
three bare coupling constants when regulated this way, as dictated by 1+1
Lorentz invariance\bpr.
One assumes that the full 3+1 Lorentz invariant theory is recovered
in the scaling region of the lattice theory for a line of tricritical
points of the coupling constants.  The tricritical points are determined
by examining 3+1 relativistic dispersion relations.

Weak coupling perturbation theory of transverse lattice non-compact
QED (TLQED) is discussed in section 4. After gauge fixing in
light-cone gauge, the UV properties of the theory are studied.
We argue that the usual diagrammatic UV divergences are cut off by the
finite transverse lattice spacing.  The transverse lattice
construction converts a four-dimensional field theory into a two-dimensional
field theory with a finite
(for finite sites on the lattice) number of ``flavors'' which is
then UV finite, diagram by diagram, for fixed lattice spacing.

In section 5 we calculate the anomalous scaling dimension of the
link fields on the lattice, and find that the interaction Hamiltonian becomes
a non-renormalizable interaction for $g^2 (a)> 4\pi$, where $g(a)$ is the
bare QED coupling constant.  The anomalous scaling dimension is
calculated by normal ordering the link fields and is non-perturbative
because the link fields are exponentials of the gauge fields.

The relationship between this phase transition
and the phase transition of
the sine-Gordon model, the quenched ladder approximation of QED,
and quenched non-compact lattice QED is discussed.  Based on these
analogies, we conjecture that this critical point
corresponds to the non-perturbative chiral symmetry breaking phase
transition in QED.

Recent interest in chiral symmetry breaking in QED was generated
by Miransky\miransky\ who used the ladder approximation
of the Schwinger-Dyson equation to argue for the existence
of a non-trivial UV renormalization group fixed point of the QED
coupling constant.  This phenomenon is
closely related to the collapse of the Dirac wavefunction in supercritical
($Z >137$, for which $\alpha=Ze^2/4\pi >1$) Coulomb fields\miransky.
The fixed point is the boundary of the chirally
symmetric ladder QED phase and its strong coupling phase which
has spontaneous chiral symmetry breaking\llb.  That the strong coupling
phase of QED breaks chiral symmetry spontaneously is understood analytically
via the strong coupling expansion of lattice gauge theory
\suss\seiler\semenoff, and via lattice gauge theory
Monte-Carlo simulations\quenched\kdk\dgroup.
It is not clear however, that lattice gauge theory
data supports the existence of a non-trivial UV fixed point for full
QED.  It may be the case that the renormalized charge of the
continuum theory vanishes at the critical point\ddgroup.

In section 6, we study the strong coupling limit of TLQED by
calculating the energy shift of the infinite coupling vacuum states
to lowest order in the inverse coupling $1/g$.  We find
that the discrete remnant of chiral symmetry on the
transverse lattice is spontaneously broken and that the chiral
condensate
$\langle \psibar \psi\rangle$ is non-vanishing for the lowest
energy state.  We discuss our results further in section 7.

\newsec{Staggered fermions for the transverse lattice}

In this section, we construct staggered fermions for the
transverse lattice, and in the process, introduce
notation for the transverse lattice construction that
will be used in later sections.

The basic strategy is to write the Dirac equation
$(i\gamma^\mu \del_\mu - m )\psi =0$ in appropriate component form, and
find a fermion equation on the transverse lattice
which reproduces these  equations in the continuum limit.
We use the chiral representation of gamma matrices
\eqn\gammas{
\gamma^0 =\left(\matrix{0&-1\cr-1&0\cr}\right)\hskip1cm
\gamma^i =\left(\matrix{0&\sigma^i\cr-\sigma^i&0\cr}\right)\hskip1cm
\gamma_5 =\left(\matrix{1&0\cr0&-1\cr}\right) \ ,
}
and define the fermion $\psi$ components
\eqn\comps{
\psi= \left(\matrix{\varphi\cr\chi\cr}\right)\hskip1cm
\varphi= \left(\matrix{\varphi^{(1)}\cr\varphi^{(2)}\cr}\right)\hskip1cm
\chi= \left(\matrix{\chi^{(1)}\cr\chi^{(2)}\cr}\right) \ ,
}
In light-cone coordinates
\eqn\nullcoor{
x^\pm = {1\over\sqrt 2}\, (x^0 \pm x^3)\hskip1cm
\del_\pm = {1\over\sqrt 2}\, (\del_0 \pm \del_3) \ ,
}
the component equations are
\eqn\compeqns{
\en{
\sqrt 2\ \del_- \chi^{(1)} =& i m \varphi^{(1)} +
 \left[ \del_1 - i\del_2 \right] \chi^{(2)} \ ,\cr
\sqrt 2\ \del_+ \chi^{(2)} =& i m \varphi^{(2)} +
\left[ \del_1 + i\del_2 \right] \chi^{(1)} \ ,\cr
\sqrt 2\ \del_+ \varphi^{(1)} =& i m \chi^{(1)} -
\left[ \del_1 - i\del_2 \right] \varphi^{(2)}\ ,\cr
\sqrt 2\ \del_- \varphi^{(2)} =& i m \chi^{(2)} -
\left[ \del_1 + i\del_2 \right] \varphi^{(1)}\ .\cr}
}

Now consider a complex one-component fermion field on a
discrete square lattice of points $\xperp= a(n_x , n_y )$, with lattice
spacing $a$ and basis vectors $\bfalpha = (a,0)$ or $(0,a)$.
In this section, the lattice is taken to be infinite.
The fermion field $\phi$ is a continuous function of the light-cone coordinates
$x^{\pm}$, and satisfies the equation
\eqn\linear{
\del_0 \phi = P_3 (\xperp ) \del_3 \phi + P_1(\xperp )\Delta_1 \phi
+ P_2(\xperp ) \Delta_2 \phi \ ,
}
where $P_1, P_2$, and $P_3$ are unknowns to be determined by matching
to the continuum equations \compeqns\ with zero mass, (adding a mass
term is more complicated and will be considered in the next section),
and $\Delta_\alpha$ is the symmetric lattice derivative
\eqn\symmetric{
\Delta_\alpha f(\xperp )={1\over 2a}\left[ f(\xperp +\bfalpha )
- f(\xperp- \bfalpha )\right] \ .
}
The fermion has the mode expansion
\eqn\mexpand{
\phi = \int d^2 k \int_{-\pi}^{\pi} d^2 \ell \ \tphi ( k^\pm, l_\alpha )\
e^{ik^+ x^-} e^{ik^- x^+} e^{i \ell_\alpha n_\alpha} \ ,
}
and in momentum space, the equation of motion is
\eqn\lineartwo{
k^0 \tphi =  k^3 {\tilde P}_3 \tphi +{\tilde P}_1
\left( {\sin \ell_1 \over a} \right) \tphi +
{\tilde P}_2 \left( {\sin \ell_2 \over a} \right) \tphi \ .
}
In the continuum limit, as $a \rightarrow 0$, finite energy states are
located about $\ell_\alpha \sim \epsilon$, or
$\ell_\alpha \sim \pi - \epsilon$.  Therefore, there are
four continuum fermion components for one transverse lattice fermion.
This is just the standard fermion doubling problem, which works to
our advantage in this case because four continuum components are desired.
Equivalently, in lattice coordinate space,
different linear combinations of four adjacent sites will correspond to four
different fields in the continuum.

To be more specific, introduce a lattice parity
$P_L [\xperp ]=(-1)^{n_x +n_y}$.  If $P_L [\xperp ]$ is $+1$ ($-1$),
then $\xperp$ is an even (odd) site.  For the moment,
consider the fermion at even or odd sites to be different continuum
fields, labeled $\phi_{\rm even}$ and $\phi_{\rm odd}$.  Making the ansatz $P_3
=P_L$,
the equation of motion \linear\ becomes
\eqn\newlinear{
\en{ \sqrt 2 \ \del_- \phi_{\rm even} =& P_1 \Delta_1 \phi_{\rm odd} + P_2
\Delta_2
\phi_{\rm odd}\ ,\cr
\sqrt 2 \ \del_+ \phi_{\rm odd} =& P_1 \Delta_1 \phi_{\rm even} + P_2 \Delta_2
\phi_{\rm even}\ ,\cr}
}
If we select $P_1 = 1$ and $P_2 = -i P_L$, then equations \newlinear\
are just the massless continuum equations for the Dirac fermion
components $\chi$ of eqn.~\compeqns.  This is not complete result,
however, because we know that there should be four continuum components.
The full result is obtained by breaking up the lattice further
into a sub-lattices graded by $(-1)^{n_y}$.  The full result is that
with the $P_1, P_2, P_3$ selected above,
\eqn\lateqns{
\en{
\chi^{(1)} =&{1\over 2}\left[ \phi (\xperp ) + \phi (\xperp + \svec )
\right] \ , \ \ \ \hbox{$\xperp$ even} \ , \cr
\chi^{(2)} =&{1\over 2}\left[ \phi (\xperp ) + \phi (\xperp + \svec )
\right] \ , \ \ \ \hbox{$\xperp$ odd} \ , \cr
\varphi^{(1)} =&{1\over 2}\left[ \phi (\xperp ) - \phi
(\xperp + \svec ) \right](-1)^{n_x} \ , \ \ \ \hbox{$\xperp$ odd} \ , \cr
\varphi^{(2)} =&{1\over 2}\left[ \phi (\xperp ) - \phi
(\xperp + \svec ) \right](-1)^{n_x} \ , \ \ \ \hbox{$\xperp$ even} \ , \cr}
}
where $\svec = a(1,1)$.  One can easily check that these fields obey the
massless version of the equations \compeqns.

Each field is associated
with the face of the lattice with center $\xperp + \half\svec$.
Label each point on the lattice by
$( (-1)^{n_x}, (-1)^{n_y} )$, so that there are four types of points
with respect to this grading.  Then $\chi^{(1)}, \varphi^{(2)}$ are
associated with type $A$ faces, and $\chi^{(2)}, \varphi^{(1)}$ are
associated with
type $B$ faces, where the faces are labeled in figure 1.

\topfigure {xsbfig1.eps} {1} {Each point on the lattice is labeled by
$( (-1)^{n_x}, (-1)^{n_y} )$.  The fields $\chi^{(1)}, \varphi^{(2)}$ are
associated with type $A$ faces, and $\chi^{(2)}, \varphi^{(1)}$ are
associated with type $B$ faces.}

Consider the symmetries of the Lagrangian
\eqn\lagrange{
\cL = i\sum_{\xperp} a^2 \ \phi^\dagger\left[ \del_0 \phi - (-1)^{n_x +
n_y}\del_3
\phi - \Delta_1 \phi + i(-1)^{n_x + n_y} \Delta_2 \phi \right]\
}
of the Dirac fermion on the transverse lattice.
In addition to $1+1$ dimensional Lorentz invariance,
it has lattice translational invariance $\xperp \rightarrow 2\bfalpha$;
as in regular lattice gauge theory, translations are shifts by
an even number of sites.  It also has the shift symmetry,
$\xperp \rightarrow \xperp + \svec$, which is interpreted as
a discrete chiral rotation on the fields, since under this transformation,
$\chi \rightarrow -\chi$ and $\varphi \rightarrow \varphi$, as is seen
by examining eqns.~\lateqns\foot{
This is denoted as a chiral rotation, although it really is a combination
of the gauge transformation $\varphi \rightarrow i\varphi$ and $\chi
\rightarrow
i\chi$, followed by the chiral rotation $\varphi \rightarrow -i\varphi$
and $\chi \rightarrow i\chi$.}.
This is also similar to the discrete chiral
symmetry found for staggered fermions on higher dimension lattices.  The
exchange symmetry of the Hamiltonian version of staggered
fermions\suss\ks\ is broken in the transverse lattice case by the
unequal treatment of the three spatial coordinates.
We are left with a discrete rotation symmetry, a rotation by $\pi$
about the $x^3$ axis, $n_\alpha \rightarrow -n_\alpha$, and a single global
gauge symmetry, $\phi \rightarrow e^{i\theta} \phi$.

A mass term for the Lagrangian should have terms of the form
${\chi^{(1)}}^\dagger  \varphi^{(1)} $ and
${\varphi^{(1)}}^\dagger  \chi^{(1)} $.  According
to the analysis which is summarized in fig.~1, the bilinear couplings will have
to be nearest neighbor, because $\chi^{(1)}$ and $\varphi^{(2)}$ live
on type $A$ faces, and $\chi^{(2)}$ and $\varphi^{(1)}$ live
on type $B$ faces.  The mass term for the Lagrangian is
\eqn\massone{
\cL^\prime_m = -m \sum_{\xperp}a^2\ \phi^\dagger (\xperp )
\phi(\xperp +P_L[\xperp ] \yhat )
}
It explicitly breaks the discrete chiral symmetry $\xperp \rightarrow
\xperp +\svec$, and leads to the correct continuum mass terms in the
equations\compeqns.
The nearest neighbor coupling in eqn.~\massone\
breaks the global $U(1)$ gauge symmetry at each site.
For each pair of sites $\xperp$ and $\xperp + P_L [\xperp ]\vec y$,
the $U(1) \times U(1)$ symmetry is broken to the diagonal
$U(1)$.  It is possible to gauge the remaining diagonal $U(1)$
symmetries, and this is discussed in ref.~\griffinb.
The construction is awkward, and we will avoid it by adding
a second flavor of lattice fermion.  Then there will exist
a mass term which preserves all of the $U(1)$ symmetry.

\newsec{Gauging transverse lattice staggered fermions}

In this section we introduce gauge fields in an attempt to
make the Lagrangian eqn.~\lagrange\ locally gauge invariant.
This however, will fail because of the 2-D gauge anomaly, and
a second set of fermion fields will have to be introduced,
leading to a fermion doubling problem in the continuum limit.

To promote $\delta_G \phi =i \Lambda \phi$ to a local gauge
symmetry, introduce the 2-D vector gauge fields $A_i$ and 2-D scalar fields
$A_\alpha$ with transformation laws
\eqn\gauging{
\en{
\delta_G A_i (\xperp, x^\pm ) =& \del_i \Lambda (\xperp, x^\pm )\ , \ \
i=0,3 \ ,\cr
\delta_G A_\alpha (\xperp, x^\pm ) =& \Delta^+_\alpha \Lambda (\xperp, x^\pm )
\ , \ \ \alpha=x,y \ . }
}
The forward lattice derivative
\eqn\forward{
\Delta^+_\alpha f(\xperp )={1\over a}\left[ f(\xperp +\bfalpha )
- f(\xperp )\right] \ ,
}
obeys the integration by parts rule
$\sum_\xperp f \Delta^+_\alpha g = - \sum_\xperp (\Delta^-_\alpha f ) g \ $,
where
\eqn\backward{
\Delta^-_\alpha f(\xperp )={1\over a}\left[ f(\xperp )
- f(\xperp - \bfalpha )\right] \ .
}
The Lagrangian for the gauge fields is
\eqn\lgauge{
\cL_{\rm gauge}= \sum_{\xperp} a^2 \left[ {1\over 4g^2_1}\left( F_{ij}\right)^2
+ {2\over 4g^2_2}\left( F_{i\alpha}\right)^2
+{1\over 4g^2_3}\left( F_{\alpha\beta}\right)^2 \right] \ ,
}
where $g_1$, $g_2$, and $g_3$ will be fixed by requiring $3+1$ Lorentz
invariance in the continuum limit.  The field strengths $F_{\mu\nu}$ are
\eqn\fstrengths{
F_{ij}=\del_i A_j - \del_j A_i \ , \ \
F_{\alpha\beta}=\Delta^+_\alpha  A_\beta - \Delta^+_\beta A_\alpha \ , \ \
F_{i\alpha}=\del_i A_\alpha - \Delta^+_\alpha A_i \ .
}
For the fermion fields, we dress the derivatives
in the Lagrangian eqn.~\lagrange\ via the minimal coupling procedure,
\eqn\covdev{
\en{
\del_i \phi \rightarrow \  D_i \phi =& \, (\del_i - i A_i)\phi \ , \cr
\Delta_\alpha \phi \rightarrow  \  D_\alpha \phi =&
\left\{ \phi (\xperp + \bfalpha )e^{-iaA_\alpha (\xperp )}
- \phi (\xperp - \bfalpha )e^{iaA_\alpha (\xperp - \bfalpha )}
\right\}/2a \ . \cr}
}
However, this construction does not yield a gauge invariant theory.  The 2-D
kinetic terms for the fermions are,
\eqn\anomaly{
\en{
\cL =& \, i \sqrt {2}\, \phi^\dagger (\del_- - i A_- )\phi +\ldots \hskip1cm
	\hbox{$\xperp$ even}\ ,\cr
=& \, i \sqrt {2}\, \phi^\dagger (\del_+ - i A_+ )\phi + \ldots \hskip1cm
\hbox{$\xperp$ odd}\ .\cr}
}
There is only a single left or right-handed fermion for each local $U(1)$
gauge symmetry, and therefore, the local $U(1)$ symmetries
are anomalous.  The anomaly breaks the $U(1) \times U(1)$ symmetry of
pairs of sites (say $\xperp$ and $\xperp + P_L[\xperp ]\vec y$) to the diagonal
$U(1)$.  The mass term eqn.~\massone\ also produced this pattern of symmetry
breaking.
In principle, one can construct transverse lattice
QED with the remaining $U(1)$ symmetry\griffinb.  However in practice, it will
be easier to add a second flavor of lattice fermions to cancel the anomalies
and
preserve the full set of $U(1)$ symmetries.  The fermion action takes the
form
\eqn\lfermi{\en{
\cL_{F} = i\sum_{f=1}^{2} \sum_{\xperp} a^2 \phi^{\dagger }_f
\biggl\{ &\left[ D_0  +(-1)^{n_x + n_y + f} D_3 \right]\phi_f\cr
-\kappa &\left[ D_x  + i(-1)^{n_x + n_y +f} D_y \right]
\phi_f\biggr\} \ , \cr}
}
where $\kappa$ is a  hopping parameter that will be fixed by requiring
3+1 Lorentz invariance.  While $\kappa =1$ in the classical continuum limit,
it will receive quantum corrections and in fact
will have to be renormalized.
With two flavors on the lattice, there will be two Dirac fermions in the
continuum limit.  Their components $\varphi$ and $\chi$
are constructed from different flavors of the lattice fermions,
i.e.~$\phi_1$ and $\phi_2$ contribute to each of the two continuum
Dirac fermions.
The components of the continuum fermions
\eqn\contfermi{
\Psi_j = \left( \matrix{ \varphi_j  \cr \chi_j  \cr} \right) \ , \ \ j=1,2
}
are
\eqn\compone{
\en{
\chi^{(1)}_1 =&{1\over 2}\left[ \phi_1 (\xperp ) + \phi_1 (\xperp + \svec )
\right] \ , \ \ \ \hbox{$\xperp$ even} \ , \cr
\chi^{(2)}_1 =&{1\over 2}\left[ \phi_1 (\xperp ) + \phi_1 (\xperp + \svec )
\right] \ , \ \ \ \hbox{$\xperp$ odd} \ , \cr
\varphi^{(1)}_1 =&{1\over 2}\left[ \phi_2 (\xperp ) - \phi_2
(\xperp + \svec ) \right](-1)^{n_x} \ , \ \ \ \hbox{$\xperp$ even} \ , \cr
\varphi^{(2)}_1 =&{1\over 2}\left[ \phi_2 (\xperp ) - \phi_2
(\xperp + \svec ) \right](-1)^{n_x} \ , \ \ \ \hbox{$\xperp$ odd} \ , \cr}
}
and
\eqn\comptwo{
\en{
\chi^{(1)}_2 =&{1\over 2}\left[ \phi_2 (\xperp ) + \phi_2 (\xperp + \svec )
\right] \ , \ \ \ \hbox{$\xperp$ odd} \ , \cr
\chi^{(2)}_2 =&{1\over 2}\left[ \phi_2 (\xperp ) + \phi_2 (\xperp + \svec )
\right] \ , \ \ \ \hbox{$\xperp$ even} \ , \cr
\varphi^{(1)}_2 =&{1\over 2}\left[ \phi_1 (\xperp ) - \phi_1
(\xperp + \svec ) \right](-1)^{n_x} \ , \ \ \ \hbox{$\xperp$ odd} \ , \cr
\varphi^{(2)}_2 =&{1\over 2}\left[ \phi_1 (\xperp ) - \phi_1
(\xperp + \svec ) \right](-1)^{n_x} \ , \ \ \ \hbox{$\xperp$ even} \ . \cr}
}
The mass term $\sum_{\xperp} a^2 {m\over \sqrt 2}\sum_j
{\overline \Psi}_j \Psi_j$ is given by
\eqn\newmass{
\cL_{m} =\sum_{\xperp} a^2 \left\{ {m\over\sqrt 2}(-1)^{n_x}
\left[ \phi^{\dagger}_1 \phi_2 + \phi^{\dagger}_2 \phi_1
\right] \right\}\ ,
}
and it preserves the $U(1)$ symmetries for all the sites.

The mass term explicitly breaks the discrete chiral symmetry
generated by $\xperp \rightarrow \xperp + \svec$.  As discussed in the
previous section, this takes $\chi_j \rightarrow -\chi_j$ and
$\varphi_j \rightarrow \varphi_j$.  This corresponds to a discrete $Z_2$
subgroup of the 4-D anomalous $U(1)$ chiral symmetry.

The 2-D gauge theory for each site on the lattice also has an
anomalous chiral transformation,
\eqn\anomtwo{
\delta \phi_f (\xperp ) = i\lambda (-1)^{f+1} \phi_f (\xperp )\ .
}
This (global in the 2-D sense) symmetry is broken at one-loop
in perturbation theory.  It corresponds in the 4-D continuum limit
to a broken axial-vector flavor symmetry, under which the continuum
components transform as
\eqn\canomtwo{
\delta \overrightarrow \Psi = i\lambda \sigma_3 \gamma_5
\overrightarrow \Psi \ , \ \
\overrightarrow \Psi = \left( \matrix{ \Psi_1  \cr \Psi_2 \cr} \right)\ ,
}
where $\sigma_3$ acts in flavor space.

The only non-anomalous continuum symmetry of this model is the
gauged $U(1)$ `total lepton number' symmetry.  There are no
global flavor symmetries for this transverse lattice model.
This is in contrast to the naive (Wilson) and staggered (Susskind)
formulations of QED on 4-D euclidean lattices\smit\kmnp.  The action for
a single 4-D naive massless fermion on the 4-D lattice has $U(4)$
vector and axial-vector flavor symmetries, which is a subgroup of the
full $U(16)$ flavor symmetries of the 16 continuum Dirac fermions of this
model.
The minimal staggered massless fermion action on the 4-D lattice has $U(1)$
vector and axial-vector flavor symmetries on the lattice, which is a
subgroup of the $U(4)$ flavor symmetries of the 4 continuum Dirac
fermions for this model.  The transverse lattice model constructed
in this section has no continuum flavor symmetries, and has only
two continuum Dirac fermions in the continuum limit.  For the
fermions on the 4-D lattices, the axial-vector flavor symmetries
which exist in the lattice action are spontaneously broken in the strong
coupling limit.  The non-vanishing of the order operator $\overline \Psi \cdot
\Psi$, which signals the breaking of the axial-vector flavor symmetries
of the lattice models, is confirmed, by Monte Carlo simulations, for the
scaling region of the theory\mpr\hp.  This order operator breaks
all of the continuum axial flavor symmetries, and one expects the
full multiplet of Goldstone bosons associated with the full set of broken
axial symmetries in the scaling regime.
In section 6, we will show that the discrete chiral symmetry of the transverse
lattice model is spontaneously broken in the strong coupling limit
by the non-vanishing vacuum expectation value of $\overline \Psi \cdot
\Psi$.

The parameter $\kappa$ is included in eqn.~\lfermi\
because 3+1 Lorentz invariance is broken down to 1+1 Lorentz invariance
by the transverse lattice construction.
One may ask whether two new parameters should really occur in the
Lagrangian, one for the $D_x$ term and one for the $D_y$ term, since
these are two separate 2-D mass terms.  The answer is no, because
there exists field redefinition that
transposes the $x$ and $y$ terms in the Lagrangian.  It is expressed as
$\phi_f\rightarrow \alpha_f\phi_f$,
where $\alpha_f$ is defined recursively,
\eqn\spindiag{
\en{
\alpha_f (\xperp ) =& i (-1)^{n_x + n_y +f} \alpha_f (\xperp - \yhat )
\ , \cr
\alpha_f (\xperp ) =& -i (-1)^{n_x + n_y +f} \alpha_f (\xperp - \xhat )
\ ,\cr
\alpha_f (0) =& 1 \cr}
}
This is a spin transformation; these transformations
are typically  applied to staggered fermion systems to diagonalize
$\gamma$ matrices in the fermion action\kmn.  Applying this particular
spin transformation to eqns.~\lfermi\ and \newmass\ interchanges the labels
$x$ and $y$.

This concludes the construction of non-compact transverse lattice QED with
staggered fermions.  The goal of the remaining sections is
to extract non-perturbative information about QED from this construction.

\newsec{Ultraviolet finiteness of perturbation theory}

In this section, the weak coupling perturbation expansion will be developed
for the transverse lattice theory with Lagrangians given by
eqns.~\lgauge\ and \lfermi.
It will be argued that
the transverse lattice regulates all the UV divergences for each
diagram in perturbation theory.
We will use this formalism in the next section to calculate the
non-perturbative scaling dimension of the interaction Hamiltonian.

Axial gauges minimize
the mixing of longitudinal and transverse degrees of freedom and are
therefore particularly useful in the context of the transverse lattice
construction.
Space-like axial gauges are problematic for weak coupling because of
difficulties implementing Gauss's law\leibbrandt, so the
light-cone gauge $A_- = 0$ will be used.  In light-cone gauge, if the field
theory is quantized on the null-plane $x^+ =0$, then $A_+$ is a constrained
field.
So in this section, we use the light-cone quantization scheme ---
light-cone gauge with the null-plane Cauchy surface.

Only half of the fermion fields $\phi^{(f)}$
satisfy dynamical equations on the null plane. With the definitions,
\eqn\fsplit{
\en{
\chi =&\phi_f \ , \ \ \ (-1)^{n_x + n_y + f} = -1 \ , \cr
\psi =& \phi_f \ , \ \ \ (-1)^{n_x + n_y + f} = +1 \ , \cr}
}
one finds that only the $\psi$ are dynamical fields.  The constraint
equations in light-cone
gauge for the fields $A_+$ and $\chi$ are
\eqn\aconst{
\del_-^2 A_+ =J_- = \left( {g_1\over g_2}\right)^2 \del_- \Delta^-_\alpha
A_\alpha- g^2_1 \sqrt {2} \psi^\dagger \psi \ ,
}
and
\eqn\cconst{
\del_{-} \chi = {1\over \sqrt 2} \left[ D_1 - iD_2 \right] \psi \ .
}
Using $\half | x^- - y^- | = 1/\del^2_-$, which satisfies $\del^2_- \, \half
|x^- - y^-|
= \delta (x^- - y^- )$, the constraint equation \aconst\ is integrated:
\eqn\intconst{
A_+ = \half \int dy^- |x^- - y^- | J_- + F x^- + G \ .
}
The constant $G(x^+ , \xperp )$ is set to zero as a gauge fixing
constraint; it fixes $x^+$ dependent (and $x^-$ independent) infinitesimal
gauge transformations.  The $F(x^+ ,\xperp )x^-$ term corresponds
to the theta angle of the Schwinger model\colemanth.  In the
continuum limit, the physical $3+1$
Lorentz covariant vacuum should correspond to $F=0$, so it can be set to
zero identically.

To remove the coupling constant dependence from the canonical commutation
relations, we let
\eqn\redefine{
A_\alpha \rightarrow g_2 A_\alpha \ .
}
The current $J_-$ and covariant derivative $D_\alpha$ must be changed
accordingly.
With this field redefinition, the light-cone momentum
$P^+ = ( P^0 + P^3 )/\sqrt 2$
and the light-cone energy $P^- = ( P^0 - P^3 )/\sqrt 2$ are
\eqn\ppm{
\en{
P^+ =& \int d\cM  \biggl\{
\del_- A_\alpha \del_- A_\alpha + i\sqrt {2} \psi^\dagger \psi
\biggr\} \ , \cr
P^- =& \int d\cM \biggl\{
{g^2_2\over 2g^2_3}F_{12} F_{12} - {1\over g_1^2 } J_- {1\over \del^2_-} J_-
+ {i\kappa^2\over \sqrt 2} ( \delta_{\alpha\beta} -
i\epsilon_{\alpha\beta} ) \psi^\dagger  D_\alpha {1\over \del_- }
\left[ D_\beta \psi \right] \, \biggr\} \ , \cr}
}
where the constraint equations have been applied, and the measure is
\eqn\measure{
d\cM = \sum_{\xperp} a^2 dx^-  \ .
}
The light-cone Hamiltonian $P^-$ can be divided into free and interacting
parts:
\eqn\hfree{
P^-_0 = \int d\cM \left\{
c_1 A_\alpha \Delta^-_\beta \Delta^+_\beta A_\alpha
+ c_2
( \Delta_\alpha A_\alpha )^2
+ {i\over \sqrt 2}\psi^\dagger \Delta_\alpha \Delta_\alpha {1\over \del_-}
\psi \right\}
}
where
\eqn\cees{
c_1 = -\half\left( g_2\over g_3 \right)^2 \ , \ \ \
c_2 = \half \left( {g^2_1\over g^4_2} - {1\over g^2_3} \right) \ ,
}
and
\eqn\hnotfree{
\en{
P^-_{\rm int} = \int d\cM \Biggl\{ &
-\sqrt 2 \left( {g_1 \over g_2 }\right)^2 \Delta^-_\alpha A_\alpha
{1\over \del_-} \left[ \psi^\dagger \psi\right]
- g^2_1 \psi^\dagger \psi {1\over \del^2_-} [ \psi^\dagger \psi ] \cr
& + {i\kappa^2 \over \sqrt 2}\psi^\dagger \biggl[ ( \delta_{\alpha\beta}-
i\epsilon_{\alpha\beta} )
D_\alpha {1\over \del_-} [ D_\beta \psi ] -
{1\over \kappa^2}\Delta_\alpha {1\over \del_-} [ \Delta_\alpha \psi ] \biggr]
\Biggr\} \ .\cr}
}

We define the beta functions
$\btl( \gamma )=a\del \gamma / \del a$ for each coupling
constant $\gamma= g_1, g_2, g_3, \kappa$.  The coupling constants $g_1$ and
$g_3$ will are fixed with respect to $g_2$ for each value of lattice spacing
$a$
by requiring that the photons obey a covariant dispersion relation.  The
appropriate renormalization scheme for covariant dispersion relations is
\eqn\rscheme{
c_1 = -\half  \ , \ \ \  c_2 = 0 \ .
}
At tree level, this implies $g_1 = g_2 = g_3$.  This relation
will receive corrections in perturbation theory; the beta functions
\eqn\betas{
\en{
\btl ( g_1 ) = \btl ( g_2 ) + \cO (g_2)\ , \cr
\btl ( g_3 ) = \btl ( g_2 ) + \cO (g_2)\ , \cr}
}
are determined by the renormalization conditions eqns.~\rscheme.

The quantum theory is defined on a square, doubly periodic, transverse
lattice
with $N^2$ sites, $N$ even.
A real scalar field  $\sigma ( \xperp ) = (\xperp + N\bfalpha )$ has the mode
expansion
\eqn\cexpand{
\sigma (\xperp ) = \sum_{\ell} \omega^{\ell\cdot n} {\tilde \sigma}(\ell )
+ \hbox{ c.c}\ ,
}
where $\omega$ is the phase factor $e^{2\pi i /N}$, the $\ell_\alpha$ are
integer momenta which take values from $- N/2$ to $N/2 -1$, and the inner
product $\ell \cdot n$ is shorthand for $\sum_{\alpha} \ell_\alpha n_\alpha$.
With these definitions, the mode expansions for the fields are
\eqn\modes{
\en{
A_\alpha =&  {1\over \sqrt {4\pi} Na} \sum_{\ell} \int_0^\infty
{d\eta \over \eta} \left\{ e^{-i\eta x^- }\omega^{\ell\cdot n} \,
a_\alpha ( \ell , \eta ) + e^{+i\eta x^- }\omega^{-\ell\cdot n} \,
a^\dagger_\alpha ( \ell , \eta ) \, \right\} \ , \cr
\psi =&  {1\over \sqrt {2\pi} Na} \sum_{\ell} \int_0^\infty
{d\eta \over \sqrt {\eta} } \left\{ e^{-i\eta x^- }\omega^{\ell\cdot n}\,
b ( \ell , \eta ) + e^{+i\eta x^- }\omega^{-\ell\cdot n} \,
d^\dagger ( \ell , \eta ) \, \right\} \ , \cr
\psi^\dagger  =&  {1\over \sqrt{2\pi} Na} \sum_{\ell} \int_0^\infty
{d\eta \over \sqrt {\eta} } \left\{ e^{-i\eta x^- }\omega^{\ell\cdot n}\,
d ( \ell , \eta ) + e^{+i\eta x^- }\omega^{-\ell\cdot n} \,
b^\dagger ( \ell , \eta )\,  \right\} \ . \cr}
}
The canonical (anti-)commutation relations for the creation and
annihilation operators are
\eqn\ccrs{
\en{
[ a_\alpha (\ell , \eta ) , a^\dagger_\beta (\ell^\prime , \eta^\prime )]
=& \delta_{\ell, \ell^\prime } \, \delta_{\alpha\beta} \, \eta \,
\delta (\eta - \eta^\prime ) \ , \cr
\{ b (\ell , \eta ) , b^\dagger (\ell^\prime , \eta^\prime ) \}
=& \delta_{\ell, \ell^\prime } \, \delta_{\alpha\beta} \, \eta \,
\delta (\eta - \eta^\prime ) \ , \cr
\{ d (\ell , \eta ) , d^\dagger (\ell^\prime , \eta^\prime ) \}
=& \delta_{\ell, \ell^\prime } \, \delta_{\alpha\beta} \, \eta \,
\delta (\eta - \eta^\prime ) \ . \cr}
}
The modes $a,b,d$ annihilate the light-cone vacuum, and the normal ordered
expressions for the fermion charge $Q_F=\sqrt 2 \int \psi^\dagger
\psi$,  momentum $P^+$, and free Hamilitonian $P^-_0$ are
\eqn\normaled{
\en{
Q_F = \sum_{\ell} \int {d\eta \over \eta} \biggl\{ &
b^\dagger(\ell , \eta ) b (\ell , \eta )
- d^\dagger(\ell , \eta ) d (\ell , \eta ) \biggr\} \ , \cr
P^+  = \sum_{\ell} \int d\eta \biggl\{ & a^\dagger_\alpha (\ell , \eta )
a_\alpha (\ell , \eta ) +  b^\dagger(\ell , \eta ) b (\ell , \eta )
+ d^\dagger(\ell , \eta ) d (\ell , \eta ) \biggr\} \ , \cr
P^-_0 = \sum_{\ell} \int {d\eta \over \eta^2} \biggl\{ & \half
k^2_{\perp} (\Delta^+ \Delta^- ; \ell ) a^\dagger_\alpha (\ell , \eta )
a_\alpha (\ell , \eta ) \cr
 + & \half k^2_{\perp} (\Delta^2 ; \ell )
\left[ b^\dagger(\ell , \eta ) b (\ell , \eta )
+ d^\dagger(\ell , \eta ) d (\ell , \eta ) \right] \, \biggr\} \ , \cr}
}
where
\eqn\kperps{
k^2_{\perp} (\Delta^+ \Delta^- ; \ell ) = \sum_\alpha
\left( { 2 \sin {\pi\ell_\alpha\over N} \over a }\right)^2
\ , \ \ \
k^2_{\perp} (\Delta^2 ; \ell ) = \sum_\alpha
\left( { \sin {2\pi\ell_\alpha\over N} \over a }\right)^2 \ .
}
The photon states $a^\dagger (\ell , \eta )\rvac$ satisfy the free-field
equation
\eqn\dispersion{
\left[\,  P^+ P^-_0 - \half k^2_{\perp} (\Delta^+ \Delta^- )\,  \right]
a^\dagger \rvac = 0 .
}
As the lattice size becomes large,
$k^2_{\perp} (\Delta^+ \Delta^- ) \rightarrow
k_1^2 + k^2_2$, where $k_\alpha = 2\pi \ell_\alpha / Na$.  Hence
eqn.~\dispersion\ is the $3+1$ Lorentz covariant free
photon dispersion relation for finite lattice size.
A similar relation holds for the fermion states.

We now argue that light-cone perturbation theory\ksop\bks\cry\ is finite,
diagram by diagram.  The $S$ matrix is
$\lstate f T \exp ( -i\int dx^+ P^{-I}_{\rm int} )\state i$, where $T$ denotes
time ordering with respect to $x^+$, and $P^{-I}_{\rm int}$ is the normal
ordered
interaction light-cone Hamiltonian \hnotfree\ in the interaction picture.
Diagrammatic perturbation
theory is generated by expanding the time ordered exponential and
inserting complete sets of intermediate states.
In general, the $S$-matrix will have an overall energy conservation factor
$-2\pi i \delta (P^-_{0,f} - P^-_{0,i} )$, and each intermediate state will
have the factor $1/(P^-_{0,f} - P^-_{0} +i\epsilon)$.  Matrix elements
of the interaction Hamiltonian with intermediate or final states
will always include the factor
$\delta (\sum_f \eta_f - \sum_i \eta_i )$, where the $\eta_f$ are
outgoing momenta and the $\eta_i$ are incoming momenta, because all vertices
conserve light-cone momentum $\eta$.  The light-cone momentum
is bounded from below by zero for all states.

One delicate aspect of light-cone perturbation theory is the limit
$\eta \rightarrow 0$ in intermediate loops.  Certain connected one-loop
diagrams
are ill defined for zero $\eta$ in continuum QED and QCD (see refs.  \pinsky\
and \thorn), and need to be regularized.  The regulator
can be removed when calculating gauge invariant combinations of one-loop
diagrams,
i.e.~the $\eta = 0$ region does not contribute to gauge invariant
processes at one-loop.
These divergences are particular to canonical
Hamilitonian perturbation theory and do not correspond to the UV divergences of
covariant perturbation theory.  Also, they are not infrared (IR) divergences
since the parity operator $P$, where $P\psi ( x^- , x^+ )P^{-1}=
\psi (x^+ , x^- )$,  acts on the modes as
$ P b^\dagger (\ell , \eta )P^{-1} \propto b^\dagger (\ell , k^2_\perp (\ell )
/2\eta )$, and interchanges the large and small $\eta$ regions.
Two popular regularization schemes for the $\eta = 0$ region are, a sharp
$\eta$ cutoff\pinsky\thorn, and the discrete light-cone approach
\casher\bpauli\mrobertson.  However, these cutoffs may not be good
regulators to higher order in perturbation theory because the $\eta = 0$ region
can contribute to connected diagrams in light-cone field
theory\griffinsg.  One signature of this problem would be the
loss of gauge or Lorentz symmetries; counterterms would have to
be added to restore the symmetries order by order in perturbation theory.

The regular UV divergences of QED arise from integration over the transverse
momentum $k_{\perp}$ of the fermions and the gauge fields
in the  $1/(P^-_{0,f} - P^- +i\epsilon)$ terms of the $S$
matrix\brodsky.
These divergences are explicitly cut off by the transverse lattice
construction, since the perpendicular momentum is bounded by $8/a^2$.
There also are IR divergences for the gauge fields and massless
fermions that arise when summing over $k_{\perp} = 0$ in the
denominators.
These correspond to the IR divergences of covariant perturbation theory,
and are regulated by introducing small mass terms: $k^2_{\perp}
\rightarrow  k^2_{\perp} + \mu^2$.

The last source of perturbative UV divergences is the
continuum 2-D field theory.  Divergent tadpoles
of the perturbation theory are eliminated by
normal ordering the light-cone Hamiltonian.
The non-local operator $1/\del_-$ in the interaction Hamiltonian
eqn.~\hnotfree\ softens the UV structure of the vertices, as opposed to
derivative interactions, which can violate UV finiteness.\colemansg.
For instance, the four fermion term in eqn.~\hnotfree\ is scaling
dimension zero (verses two) because of the non-local $1/\del_-^2$
factor.  And by further power counting arguments, the interaction light-cone
Hamiltonian is UV finite, diagram by diagram.

In principle, the 2-D fermions
$\psi$ in the light-cone Hamiltonian can be bosonized.  With the bosonization
relations $\psi = :\exp {i\sqrt {4\pi} \Phi}:$ and $\psi^\dagger = :\exp
{-i\sqrt {4\pi} \Phi}:$, where $\Phi$ is a canonical boson,
the light-cone Hamiltonian of TLQED
is mapped to a bosonic light-cone Hamiltonian
with non-derivative interactions.  It
is well known that a bosonic theory in two dimensions with no derivative
interactions is UV finite, diagram by diagram\colemansg.  It
is also possible to bosonize the 2-D covariant Lagrangian.  Then
the bosonization dictionary which translates between fermions and
bosons will be more complicated\mandelstam.

\newsec{The non-perturbative ultraviolet divergence at $g^2_2 = 4\pi$}

While the transverse lattice theory of QED is UV finite diagram
by diagram, it can happen that an infinite number of diagrams
conspire to generate a new UV divergence.  This phenomenon occurs in the 2-D
sine-Gordon model\colemansg\samuel\mrg\agg.  The basic signature of this
phenomenon in the sine-Gordon model is that the anomalous scaling dimension of
the interaction $( \alpha /\beta^2 ) \cos (\beta \phi)$ is greater than two
for $\beta^2 > 8\pi$, and the interaction is becomes non-renormalizable.
For this region of coupling, the energy density is unbounded from below
\colemansg, and the connected Green's functions diverge order by order in
$\alpha$, starting at order $\alpha^2$\samuel\mrg.

For TLQED we will now calculate the leading anomalous scaling dimension of the
interaction light-cone Hamiltonian \hnotfree.   It is obtained by
considering the parts of eqn.~\hnotfree\ that contain non-interacting products
of link fields.  The prototypical term of this type is
\eqn\asint{
{\kappa^2_0\over a^2_3} \psi^\dagger \epsilon_{\alpha\beta} D_\alpha
{1\over \del_-} \left[ D_\beta \psi \right] \ ,
}
where $\kappa^2_0$ is the bare coupling and $a_3$ is the cutoff of the
2-D continuum theory.  This is a bare expression, since it depends
upon $a_3$, and it needs to be renormalized with respect to an arbitrary mass
scale.  We will calculate the divergent tadpole contributions and
renormalize this term.
In eqn.~\asint, the factor $1/a^2_3$ accounts for the naive scaling dimension
of this interaction, which is $\half + \half -1=0$, where each $\half$ comes
from the fermions and $-1$ comes from $1/\del_-$.  Since the fermion fields
$\psi^\dagger$ and $\psi$ in eqn.~\asint\  occur
at different lattice sites and therefore anticommute, and
the two link fields commute because of $\epsilon_{\alpha\beta}$,
we only have to normal order each
link field to obtain the tadpole contributions.  Consider the
exponential
\eqn\expo{
e^{iag_2A_\alpha} = :e^{iag_2 A_\alpha}:e^{+\half (g_2 a)^2
[A_\alpha^+ , A_\alpha^- ]} \ ,
}
where $A^+$ ($A^-$) includes only raising (lowering) operators in the
fields mode expansion \modes.  After applying the commutation relations,
we get
\eqn\acomm{
[A_\alpha^+ , A_\alpha^- ]={1\over 4\pi (Na)^2} \sum_{\ell}
\int^{\Lambda^+_{\ell} }_{\delta^+_{\ell} } {d\eta \over \eta} \ .
}
The small $\eta$ regulator  $\delta^+_{\ell}$ and the large $\eta$ regulator
$\Lambda^+_{\ell}$ are related by $x_3$ parity, as discussed in the previous
section and in ref.~\griffinsg.  The relationship is
\eqn\regs{
\delta^+_{\ell} = { k_{\perp}^2 (\Delta^+ \Delta^- ; \ell )\over 2
\Lambda^+_{\ell} } \ .
}
In terms of a fixed $x_3$ momentum cutoff $\Lambda \approx 1/a_3$,
the large $\eta$ cutoff $\Lambda^+_\ell$ is given by the 2-D relativistically
correct expression,
\eqn\moreregs{
\Lambda^+_{\ell} = {\Lambda + \sqrt {\Lambda^2
+ k_\perp^2 (\ell ) } \over \sqrt 2}\,  \ .
}
Here, $k_\perp$ plays the role of a mass for each 2-D theory.
In the limit $\Lambda >> k_\perp (\ell )$,
\eqn\logint{
\int^{\Lambda^+_{\ell} }_{\delta^+_{\ell} } {d\eta \over \eta} =
\ln \left[ { 4 \Lambda^2 \over k_{\perp}^2 } \, \right] \ ,
}
and
\eqn\expotwo{
e^{iag_2A_\alpha} = :e^{iag_2 A_\alpha}:
\left\{ { \left[ \prod_{\ell} k^2_{\perp} (\ell ) \right]^{1/N^2}
\over 4\Lambda^2} \right\}^{g^2_2 / 8\pi} \ .
}
In eqns.~\logint\ and \expotwo, the IR divergence at $\ell = 0$ is regulated
by adding a small mass: $k_\perp (\ell = 0 ) \rightarrow \mu^2$.
We see that the exponentials have anomalous scaling dimension $g^2_2 /4\pi$,
i.e.~they scale as $\Lambda^{-g^2_2 /4\pi}$, where $\Lambda$ is the
UV momentum cutoff.
The interaction term \asint\ is multiplicatively renormalized by defining
a renormalized coupling $\kappa (m)$ as
\eqn\mrenorm{
\kappa_0^2 = Z_{\kappa} \kappa^2 (m) \ , \ \ \
Z_{\kappa} = {1\over 4}(2ma_3 )^{2-g^2_2 / 2\pi} \
 \prod_{\ell} \left( k_{\perp} (\ell )\right)^{g^2_2 / 4\pi N^2}\ ,
}
where $m$ is an arbitrary mass scale.
The renormalized interaction term is then
\eqn\newasint{
m^{2-g^2_2 /2\pi} \kappa^2 (m)
\psi^\dagger \epsilon_{\alpha\beta} D_\alpha
{1\over \del_-} \left[ D_\beta \psi \right] \ .
}
For $g^2_2 <  4\pi$, the interaction
term has dimension less than two.
For this region of coupling constant, the UV finiteness of each
diagram in the theory is sufficient to guarantee finiteness of the
full theory.  For $g^2_2 = 4\pi$ the interaction term \asint\
is a marginal operator, and the theory will be well defined
if the renormalization of $g_2$ with respect to the
2-D continuum theory is allowed.  This is the situation for the
sine-Gordon model at its critical point\miransky\agg.
For $g^2_2 > 4\pi$ the theory is non-renormalizable,
the hopping parameter $\kappa$ has negative scaling dimension,
and the operator product of the interaction Hamiltonian with
itself is too singular to allow consistent perturbation theory
about the free-field vacuum.

Therefore, for TLQED, we find the somewhat surprising result that the weak
perturbation theory is valid only for $\alpha (a) = g^2_2 /4\pi <1$,
independent of $a$.  The coupling $g^2_2 (a)$ is the bare coupling
and in the scaling regime of full TLQED it may be quite
far from the renormalized QED coupling constant $g_{\rm ren}$.
Only for very weak coupling is $g_2 (a) \approx g_{\rm ren}$ in the full
theory.
However, recall that in the quenched approximation of lattice QED\quenched,
chiral symmetry is spontaneously broken beyond a certain critical value $\alpha
\sim 1$.  Similarly, the analytic calculations in the ladder approximation of
quenched QED also exhibit a critical coupling which corresponds
to the chiral symmetry breaking phase transition\miransky\llb.  We
therefore make the conjecture that $g^2_2 (a) =4 \pi$ is in general the
chiral symmetry breaking critical point in TLQED, and that specifically, in the
quenched approximation of TLQED,
for which $g_2 (a) = g_{\rm ren}$, chiral symmetry is broken for $\alpha >1$.
This is discussed further in section 7.

\newsec{Strong coupling limit}

Does TLQED realize spontaneous chiral symmetry breaking in the strong coupling
regime?  This means that a non-vanishing chiral condensate $\langle
\overline \Psi \cdot \Psi\rangle$
must appear, or equivalently in terms of the the lattice fermions,
\eqn\condensate{
\sum_{\xperp}(-1)^{n_x}
\lstate {\rm vac} \phi^{\dagger}_1 \phi_2 + \phi^{\dagger}_2 \phi_1
\state {\rm vac} \ \neq 0 \ ,
}
where $\state {\rm vac}$ is the full interacting vacuum state.  Such a
non-vanishing
condensate would signal the spontaneous breaking of the discrete
chiral symmetry of the lattice theory.  Since it is a
discrete symmetry in the strong coupling region, there will be no
accompanying Goldstone boson in this region, and
Coleman's theorem\colthm, prohibiting  spontaneous breaking
of continuous internal symmetries in two dimensions without anomalies
or a Higgs mechanism, will not be violated.
The discrete chiral symmetry of the lattice model corresponds to
the 4-D anomalous $U(1)$ chiral symmetry, and we would not
expect Goldstone bosons for this broken symmetry in the scaling regime
of the transverse lattice model.
However, non-vanishing of the condensate
eqn.~\condensate\ in the scaling regime would also signal the breaking of the
non-anomalous continuum $U(2)$ axial flavor symmetries, and we would expect
their accompanying Goldstone bosons in the scaling regime.

We will now show that spontaneous chiral symmetry breaking does
occur in TLQED in the infinite coupling
$g_i\rightarrow \infty$ limit, where $i=1,2,3$. (Here we assume
$g_1 \sim g_2 \sim g_3$.),  by calculating the
energy difference between various vacuum configurations defined below
to lowest order in $1/g$.
As we will see, this
calculation is complicated by the fact that the field theory of
rigid rotators is fraught with divergences.  In the end however, the vacuum
energy density shift will be a finite quantity.

Unlike the previous weak coupling analysis, it is
convenient to perform the analysis in the $A_3 = 0$ gauge,
and with equal time quantization.
The Hamiltonian density is then
\eqn\newham{
\cH = {1\over 2}\left( {g_2\over a}\right)^2 E^2_\alpha + A_0 \left[
\Delta^-_\alpha E_\alpha + a^2 j_F \right] + H_F + \cO ({1\over g^2}) \ ,
}
where $E_\alpha$ is the electric field and momentum conjugate to
$A_\alpha$, and $j_F = \sum_f \phi^{\dagger (f)} \phi^{(f)}$ is the
fermion current.  Gauss's law,
\eqn\gauss{
\cG (\xperp )= \Delta^-_\alpha E_\alpha + a^2 j_F (\xperp ) = 0 \ ,
}
is obtained by integrating
out $A_0$, and is treated in the quantum theory
as the weak constraint $\langle \cG_\alpha \rangle = 0$ for all
physical correlation functions.  To leading order in $g$, the vacuum
must satisfy
\eqn\hwstate{
E_\alpha (\xperp ) \rvac = 0 \ , \ \forall \xperp \ .
}
The system will be quantized with respect to this `free-field' vacuum.
The condition that all modes of canonical momentum annihilate the
vacuum is reminiscent of the rigid rotator in quantum mechanics.

To regulate the IR behavior of the system, introduce periodic
boundary conditions in the continuous spatial direction $z=x_3$,
\eqn\pbc{
-L \leq z \leq L \ .
}
The mode expansions for the second quantized fields are
\eqn\expand{
\en{
E_\alpha =& {1\over 2L}\left\{ \sum_{n=1}^{\infty}\left[
E^n_\alpha  e^{+i\pi n z/L} + E^{*n}_\alpha  e^{-i\pi n z/L} \right]
+ E^0_\alpha \right\} \ ,\cr
A_\alpha =& \sum_{n=1}^{\infty}\left[
A^n_\alpha  e^{+i\pi n z/L} + A^{*n}_\alpha  e^{-i\pi n z/L} \right]
+ A^0_\alpha  \ , \cr}
}
where $E^{*n}_\alpha$ ($A^{*n}_\alpha$) are the complex conjugates of
$E^{n}_\alpha$  ($A^{n}_\alpha$), and {\it not} hermitian conjugates in the
sense of raising and lowering operators of the harmonic oscillator.
The canonical commutation relations in terms of the modes are
\eqn\ccr{
\en{
[ A^{*n}_\alpha (\xperp) , E^m_\beta (\yperp)] =&
i\delta_{\alpha \beta}\delta^{n+m}\delta_{\xperp, \yperp} \ , \cr
[ A^{n}_\alpha (\xperp) , E^{*m}_\beta (\yperp)] =&
i\delta_{\alpha \beta}\delta^{n+m}\delta_{\xperp, \yperp} \ , \cr}
}
The free Hamiltonian and momentum for the gauge fields are given by
\eqn\modeham{
\en{
H^0_{\rm gauge}=& {1\over 2L}\left({g_2\over a}\right)^2 \sum_{\xperp,\alpha}
\left[ \
\sum_{n=1}^{\infty} E^n_\alpha E^{*n}_\alpha + \half ( E^0_\alpha )^2 \
\right]  \ , \cr
P_{\rm gauge}=& \sum_{\xperp,\alpha,n} {in\pi\over L}\left[ A^{*n}_\alpha
 E^{n}_\alpha - A^{n}_\alpha E^{*n}_\alpha \right] \ . \cr }
}
All $E^n_\alpha$ and $E^{*n}_\alpha$ are lowering
operators and annihilate the free-field vacuum,
and the modes $E^n_\alpha , A^{*n}_\alpha$
( $E^{*n}_\alpha , A^{n}_\alpha$) are eigenstates of momentum
$P$ with eigenvalues $n$ ($-n$).

Creation operators in the Hilbert space are exponentials of the
modes  $A^n_\alpha $ with charge $n$.
For instance, the momentum zero mode  state $e^{in a A^0_\alpha}\rvac$ has
energy eigenvalue $(g_2 n)^2 / 4L$.  Each state in the Hilbert space
corresponds to a wavefunction in a first quantized theory where the
dimensionless quantity $a A^n_\alpha$ plays the role of a coordinate.
This relation can be used to calculate correlation functions.
The correlation function of states is non-vanishing
only if the total charge in the exponentials of the wavefunctions vanish.

More precisely, for the zero mode expectation values, the two point correlator
is
\eqn\twopoint{
\langle e^{-imaA^0_\alpha}\  e^{inA^0_\beta} \rangle \equiv
\cN^{-1}\delta_{\alpha\beta} \int^\infty_{-\infty} dx e^{i(n-m)x}
= \delta_{\alpha\beta} \cN^{-1} \delta (n-m) \ .
}
Note that for the compact $U(1)$ theory, the integration region for
$x$ would be $[-\pi, \pi ]$, $n,m$ would be integers, and the correlator
eqn.~\twopoint\ would be $\delta_{n,m}$.  In the non-compact case at hand,
the result is a normalized Dirac delta function, which is ill-defined
for arbitrary real $n,m$; only for ``integer'' $n,m$ do the non-compact and
compact results coincide\balaban.
In this section, we will evaluate such correlators with non-integer
arguments, and regulate the result by defining the cutoff delta function
\eqn\deltaone{
\delta_\Lambda (x)= {1\over 2\pi} \int^{\Lambda}_{-\Lambda}dk e^{ikx}
}
The normalization is given by $\cN^{-1} = \delta_\Lambda (0)$.

This is not the only expression which needs to be regulated in the
theory.  Consider the exponential of the field $e^{ia A_\alpha (z)}$ acting on
the vacuum.  This expression appears in the interaction Hamiltonian;
it represents
a link field carrying flux from $\xperp$ to
$\xperp +\bfalpha $ and has energy eigenvalue
\eqn\bigev{
H_{\rm gauge}^0 \ e^{ia A_\alpha (z)} \rvac =
{g_2^2\over 4L} \left[ 1 + 2\sum_{n=1}^{\infty} \right]
\ e^{ia A_\alpha (z)} \rvac = {g_2^2\over 2} \delta (0) e^{ia A_\alpha (z)}
\rvac
}
The energy is infinite because the exponential is a product of an infinite
number of states.  The exponential receives contributions from
all of the `standing waves' $A_\alpha^n$ and $A_\alpha^{*n}$ in the box.
To regulate this UV divergence, introduce a cutoff in the number of modes
counted in the delta function
\eqn\deltatwo{
\delta_{\Lambda^\prime}^L (z)= {1\over 2L}
\sum_{j=-\Lambda^\prime}^{\Lambda^\prime} e^{i\pi jz/L} \ .
}
Then the energy of the exponential is ${g_2^2\over 2}
\delta_{\Lambda^\prime}^L (0)$.  The energy is proportional
to the number of links and to the square of the flux carried
by each link.

The next to leading order contribution to the
Hamiltonian comes from the fermions and their interactions with
the gauge field.
We adopt the equal time anticommutation relations for the fermions
\eqn\fermiccr{
\{\, \phi^{\dagger}_f(z,\xperp ) \, , \, \phi_g(z^\prime, \yperp ) \, \} =
{1\over a^2} \delta (z-z^\prime )\delta_{\xperp , \yperp } \delta_{fg} \ .
}
The free-field Hamiltonian density for the fermions is
\eqn\fham{
\cH^0_F = -i\sum_{\xperp,f} a^2(-1)^{n_x + n_y + f} \phi^{\dagger}_f
\del_z \phi_f \ .
}
Because of the minus signs in this expression, the mode expansion for the
fermions is
\eqn\fmodes{
\en{
\phi_f =& {1\over \sqrt {2La^2}}\left[ \sum_{n\neq 0}
\left( b^{\dagger (f)}_n e^{-i\pi n z/L} + d_n^{(f)}
e^{i\pi n z/L} \right) + b_0^{(f)}
\right]\, ,  (-1)^{n_x + n_y +f} = +1 \, , \cr
\phi_f =& {1\over \sqrt {2La^2}}\left[ \sum_{n\neq 0}
\left( b_n^{(f)} e^{-i\pi n z/L} + d^{\dagger (f)}_n
 e^{i\pi n z/L} \right) + b^{\dagger (f)}_0
\right]\, ,  (-1)^{n_x + n_y +f} = -1 \, , \cr}
}
where for each site $\xperp$ and flavor $f$,
\eqn\modeaccr{
\{ b_n , b_m^\dagger \} = \delta_{n,m} \ , \hskip1cm
\{ d_n , d_m^\dagger \} = \delta_{n,m} \ , \hskip1cm
\{ b_0 , b_0^\dagger \} = 1 \ .
}
The normal ordered free-field Hamiltonian is given by
\eqn\noham{
H^0_F = \sum_{\xperp,f,n} {n\pi\over L} \left( b^{\dagger (f)}_n
b^{(f)}_n +  d^{\dagger (f)}_n  d^{(f)}_n \right) \ ,
}
and
\eqn\fno{
\lvac b^\dagger_n =\lvac d^\dagger_n = b_n \rvac = d_n \rvac = 0 \ ,
\ \ \ n>0 \ ,
}
for each fermion flavor.
The zero modes appear in the charge operator
\eqn\fcharge{
Q (\xperp )  = \int dz j_F = \half [ b^{\dagger (f)}_0 , b^{(f)}_0 ]
+ \ldots \ ,
}
and in the mass operator
\eqn\fmass{
M (\xperp ) = a^2 \int dz \left[ \phi^{\dagger}_1 \phi^{\ }_2 +
\phi^{\dagger}_2
\phi^{\ }_1 \right] = b^{\dagger (1)}_0 b^{(2)}_0  +
b^{\dagger (2)}_0 b^{(1)}_0   + \ldots \ .
}
The chiral condensate order parameter is
proportional to $\sum_{\xperp} (-1)^{n_x} M(\xperp )$.

The vacuum states of the full theory to order $\cO (g^0 )$ will be a
direct product of the gauge field vacuum $\state 0$ and the highest weight
states for the fermion zero modes.
To discuss chiral symmetry breaking in the zero mode sector of the fermion
theory, we diagonalize
the charge $Q$ and mass operator $M$
simultaneously, via the Bogolubov transformation
\eqn\bogo{
\en{
b^{(1)}_0 =& {1\over \sqrt 2}\, ( a_0 + i c_0 )  \hskip1cm
b^{\dagger (1)}_0 = {1\over \sqrt 2}\, ( a_0^\dagger - i c_0^\dagger )
\ , \cr
b^{(2)}_0 =&{1\over \sqrt 2}\, ( a_0 - i c_0 ) \hskip1cm
b^{\dagger (2)}_0 = {1\over \sqrt 2}\,  ( a_0^\dagger + i c_0^\dagger )
\ , \cr}
}
where $\{ a^{\dagger}_0 , a_0 \} = \{ c^{\dagger}_0 , c_0 \} =1$.
Then the mass and charge operators in the zero mode sector for each site are
\eqn\mcharge{
M= \half [a^{\dagger}_0 , a_0] - \half [c^{\dagger}_0 , c_0 ]\ ,\hskip1cm
Q= \half [a^{\dagger}_0 , a_0] + \half [c^{\dagger}_0 , c_0 ] \ .
}
The operators $a_0^\dagger , a_0$ and $c_0^\dagger , c_0$ act on
two level systems.  The $a$ operators are
raising and lowering operators for the states ${\state \uparrow}_a$ and
${\state \downarrow}_a$,
\eqn\twolevel{
a^{\dagger}_0 {\state \downarrow}_a = {\state \uparrow}_a, \hskip.5cm
a^{\dagger}_0 {\state \uparrow}_a = 0 , \hskip.5cm
a_0 {\state \uparrow}_a = {\state \downarrow}_a, \hskip.5cm
a_0 {\state \downarrow}_a = 0\ .}
The vacuum states in the fermion sector are direct products of the
two level states in the $a$ and $c$ systems,
\eqn\fvacuum{
\state {+} = {\state \uparrow}_a {\state \downarrow}_c \ ,\hskip1cm
\state {-} = {\state \downarrow}_a {\state \uparrow}_c \ .}
They satisfy $Q \state {\pm} = 0$ (Gauss's law) and
$M \state {\pm} = \pm \state {\pm}$.
The vacuum for each site on the lattice is therefore doubly degenerate
at $\cO (g^0 )$.  Note that fermion zero mode expectation values vanish:
$\langle b_0^{(f)} \rangle = 0$ and $\langle b_0^{\dagger(f)} \rangle =0$.
The non-vanishing two point functions are
\eqn\ftwopoint{
\en{
\langle\,  b^{\dagger (f)}_0 b^{(f^\prime)}_0 \, \rangle =&
+  \langle\,  b^{(f)}_0 b^{\dagger (f^\prime)}_0 \, \rangle = \half \ ,
\ \ \ \ \ f=f^\prime \ ,\cr
\langle\,  b^{\dagger (f)}_0 b^{(f^\prime)}_0 \, \rangle =&
- \langle \, b^{(f)}_0 b^{\dagger (f^\prime)}_0 \, \rangle = \half M \ ,
\ \  f\neq f^\prime \ . \cr}
}

We now show that
the degeneracy of the vacuum state is broken in perturbation theory by the
interaction Hamiltonian
\eqn\hint{
H_{\rm int} = i\kappa \sum_{\xperp,f}a^2 \int dz \phi^{\dagger}_f \left[
D_x  - (-1)^{n_x + n_y + f}D_y \right] \phi_f \ ,
}
which is a gauge invariant operator since $[ \cG (\xperp ) ,H_{\rm int}]=0$.
In the context of the 4-D transverse lattice theory,
the constant $\kappa$ is dimensionless, since the fermions are dimension
$3/2$ and the lattice derivatives go like $1/a$ and are dimension $1$.
When power counting for the continuous 2-D theory however, the
fermions are dimension $1/2$ and the derivative is dimension $0$.  Therefore
$\kappa$ is dimension $1$ in the context of the 2-D field theory:
$\kappa \sim a/a_3$, where $a_3$ is the UV cutoff of the 2-D
theory at each site.  To regulate the energy of states, we have introduced
a cutoff in the number of modes, $\delta^L_{\Lambda^\prime} (0)$.  The UV
cutoff $a_3$ is given by $a_3 \sim 1/\delta^L_{\Lambda^\prime} (0)$,
so that
\eqn\newkappa{
\kappa = \kappa^{\prime} a\delta^L_{\Lambda^\prime} (0)  \ ,
}
where $\kappa^\prime$ is a scale independent constant.

The first order shift $\langle H_{\rm int} \rangle$ in the vacuum energy
vanishes because the expectation value of a single link field vanishes.  The
second order shift is given by
\eqn\etwo{
W_2 = {\sum_{n}}^\prime \ {\lvac H_{\rm int}\state n \, \lstate n H_{\rm int}
\rvac
\over 0 - W_n } \ ,
}
where $W_n ={g_2^2\over 4L} \delta_{\Lambda^\prime}^L (0) + W_{n,F}$
is the energy eigenvalue of link states $\state n$, and
$W_{n,F}$ is the fermion sector contribution.  We will calculate the
shift in the vacuum energy due to the assignment of the fermion vacuum
to the zero mode states $\state \pm$ at each site on the lattice, which
will be denoted as $\delta W_2$.

To calculate the second order energy shift of the vacuum,
we need the correlation function
\eqn\twocorr{
\int dz f(z) \int dz^\prime g(z^\prime ) \
\langle e^{-iqaA_\alpha (z)} e^{+iq^\prime aA_\beta (z^\prime )}\rangle
\ .
}
This correlator occurs when summing over intermediate states in
eqn.~\etwo.
Integrating out the zero modes $A^0_\alpha$ in the exponentials yields the
factor
$\delta_{\alpha\beta}\delta_\Lambda (q-q^\prime ) / \delta_\Lambda (0)$.
{}From the next lowest mode $i(A^1_\alpha - A^{*1}_\alpha )$, there
is the factor
\eqn\sinfactor{
\en{
\delta_\Lambda & (\, 2\sin ( z\pi /L ) -
2\sin ( z^\prime \pi /L ) \, ) / \delta_\Lambda (0)\cr
= &{L\over 2\pi}\left[  \delta_\Lambda (z-z^\prime )
+\Theta (z^\prime ) \delta_\Lambda (z+z^\prime -L )
+\Theta (-z^\prime) \delta_\Lambda (z+z^\prime +L ) \right]/ \delta_\Lambda (0)
\ .\cr}
}
The only term on the r.h.s.~of eqn.~\sinfactor\ that contributes to the
correlator is $\delta_\Lambda (z-z^\prime )$.  The other two terms
will lead to vanishing contributions because there is no overlap with
these delta functions and the delta functions that appear when integrating out
the cosine terms; for instance, integrating out $(A^1_\alpha + A^{*1}_\alpha )$
yields a term $\delta_\Lambda  (\, 2\cos ( z\pi /L ) -
2\cos ( z^\prime \pi /L ) \, )$ which has no overlap with the second
two terms in eqn.~\sinfactor.  In the presence of the first term of
eqn.~\sinfactor,
all the other modes in the correlator contribute factors of unity.
Therefore, the correlator eqn.~\twocorr\ is given by
\eqn\twoanswer{
\delta_{\alpha\beta}{ \delta_{\Lambda}(q-q^\prime )\over \delta^2_\Lambda (0)}
\int dz f(z) g(z)
}

The parameter $\Lambda$ is an ultraviolet regulator.
If the $z$ direction were discretized, then the $\sin (n z \pi /L)$
and $\cos (mz\pi /L )$ terms which appear as arguments in the delta
functions of the correlation function \twocorr\ would take on
discrete values ($n,m$ would be bounded by $\sim[ 2\pi /a_3 ]$, where
$a_3$ is the lattice spacing in the discretized $z$ direction).
For integer charges $q,q^\prime$, which is all that we will have
to consider in this section, all of the correlation functions would
then be normalizable.  The discrete version of the normalized delta
function $\delta_{\Lambda} (z) / \delta_{\Lambda} (0)$ would be
$(a_3 )^{-1}\delta_{z_i} / (a_3 )^{-1}$.  Hence
$\delta_\Lambda (0)\approx 1/a_3$

While eqn.~\etwo\ is a complicated sum over four point correlation functions
of the
fermion modes, the only the terms which
contribute to the shift in vacuum energy $\delta W_2$ are products of
four fermion zero modes.
The non-trivial part of this observation is that a typical
two zero mode contribution $\langle b^{(f)}_0 b^{(f)}_n
b^{\dagger (f^\prime)}_n b^{\dagger (f^\prime )}_0 \rangle$ is
proportional to $\langle b^{(f)}_0  b^{\dagger (f^\prime )}_0
\rangle \delta^{ff^\prime}$, and this by the first of eqns.~\ftwopoint\ is
independent of the choice $\state +$ or $\state -$ for the vacuum state at
that site.
Using the link field two point correlation function \twocorr\ given by
eqn.~\twoanswer\
and the fermion zero mode two point correlators eqns.~\ftwopoint,
the shift in the energy density is
\eqn\dwtwo{
\delta  w_2 = { {\kappa^\prime }^2 \over 16\pi g_2^2 }
\left[ 1 \over
(La)^2  \right]
\sum_{\xperp}  M(\xperp ) \left[ M(\xperp + \xhat )
-  M(\xperp + \yhat ) \right]\,  \ .
}
This is minimized for $M(\xperp ) M(\xperp +\xhat ) = -1$ and
$M(\xperp ) M(\xperp +\yhat ) = +1$.  There are two
fermion vacuum configurations, related by an overall
sign change, that obey these conditions and the
symmetry of these ground states is made clear by figure~2.
Both configurations break the discrete $U(1)$ axial chiral symmetry since
the order parameter $\sum_{\xperp} (-1)^{n_x} M(\xperp )$ is
non-vanishing for these vacuum configurations.  If the order parameter
is non vanishing in the scaling regime, then the full set of
non-anomalous continuum axial flavor symmetries will be broken.

\topfigure {xsbfig2.eps} {2} {The plus and minus signs for each site
refer to the fermion zero mode states $\state +$ and $\state -$.
Up to an overall change in sign, this configuration minimizes
the order $1/g^2$ correction to the vacuum energy density in the strong
coupling limit.}

There is a simple way of approaching the continuum limit of this
leading order result in the strong coupling regime such that
eqn.~\dwtwo\ remains finite, i.e.~let the
longitudinal IR regulator $L\rightarrow \infty$ and the
transverse UV regular $a\rightarrow 0$ such that $La$ remains finite.
So although the energy and correlators of link fields require
UV regulators, the shift in the vacuum energy density is finite
in the continuum limit.  We briefly list the
sources of the regulated divergences that contribute to eqn.~\dwtwo.
The product of four zero modes contributes $1/L^2 a^4$, the energy in the
denominator
of \etwo\ contributes $1/\delta^L_{\Lambda^\prime} (0)\sim a_3$, from the
integral over intermediate states we get $1/\delta_\Lambda (0) \sim a_3$,
from the $\kappa^2$ coupling constant there is a factor of $a^2 /a^2_3$,
examination of eqn.~\hint\ shows that $H_{\rm int}^2$ contributes a factor
of $a^2$,
and we multiply by $1/a^2$ to make \etwo\ into a density.  The result
is the net factor of $1/(La)^2$.

To interpret this result further,
consider the spin transformation $b^{(2)}_0 \rightarrow \alpha b^{(2)}_0$,
where $\alpha (\xperp ) = (-1)^{n_y}$.  Following the analysis
of Semenoff\semenoff, define the vector
\eqn\psidef{
\psi = \left( \matrix{ b^{(1)}_0  \cr b^{(2)}_0  \cr} \right) \ ,
}
and the currents $S_j = \psi^\dagger \sigma_j \psi$ where $\sigma_j$ are
Pauli matrices. Then the Hamiltonian density in the zero mode sector
that has expectation value given by eqn.~\dwtwo\ can be written as
\eqn\spinham{
\delta  e_2 = { {\kappa^\prime }^2 \over 16\pi g_2^2 }
\left[ {1 \over
(La)^2 } \right]
\sum_{\xperp} {\vec S}(\xperp )\cdot \left[{\vec S}(\xperp + \xhat )
+  {\vec S}(\xperp + \yhat ) \right]\, + \hbox{const.} \ .
}
This is the Hamiltonian density for the quantum spin $\half$ Heisenberg
antiferromagnet,
and the configuration given by fig.~2 is just the classical ground
state of the system\lmattis.  It has N\'eel order, i.e. the expectation
value of eqn.~\dwtwo\ is non-vanishing and the global flavor $SU(2)$
of the Hamiltonian \spinham\ is spontaneously broken.
We can consider  eqn.~\spinham\ to be Hamiltonian in the
fermion sector to leading order in the strong coupling expansion.

To study chiral symmetry breaking to higher order in the strong coupling
expansion, we need to treat the quantum fluctuations of the spin
$\half$ Heisenberg antiferromagnet in the zero mode sector,
and include the effect of non-zero modes on the vacuum state.
There is no exact solution
of the ground state of the quantum d=2 quantum spin $\half$ Heisenberg
antiferromagnet\kls, and no proof that N\'eel order persists in the full
quantum theory.  However, numerical simulations indicate that this may be the
case\gss.  For a similar analysis of regular
Hamiltonian lattice gauge theory the situation is better, because
N\'eel order has been proven to exist in three dimensions\semenoff\kls.

\newsec{Discussion}

The transverse lattice regulation of QED that has been studied
in this paper is a `minimal' way of regulating the
diagrammatic divergences of the perturbation theory,
and it exhibits a phase transition at a critical value of the
lattice QED coupling constant, and chiral symmetry breaking
in the strong coupling regime\foot{If one formulates QED with one lattice
and three continuum dimensions, then the diagrammatic divergences
will not be regulated by the lattice, and chiral symmetry breaking
will not appear in the strong coupling expansion of the lattice theory.
This is shown by choosing a gauge where the lattice gauge field is set
to zero.}.

In section 5, we took advantage
of the UV finiteness of each diagram in weak perturbation theory
to find a non-perturbative UV divergence at $g^2_2 (a) = 4\pi$.
The transverse lattice regulates the usual UV divergences of four
dimensional QED, but the `finite' two-dimensional field theories
for each site conspire to generate a non-renormalizable interaction.
The signature of the non-renormalizability is the anomalous scaling
dimension of the interaction Hamiltonian.  If the dimension of any part of the
interaction Hamiltonian is greater than two, then the perturbation theory
about the free-field vacuum will be ill defined.
One can calculate the anomalous dimension of the interaction
Hamiltonian because the coupling constant $g_2 (a)$ is not
renormalized in the 2-D continuum perturbation theory for $g_2^2 < 4\pi$.
Note that there is no plaquette term in the interaction Hamiltonian,
since we have studied non-compact QED,
which would have a higher scaling dimension than the term we
considered\foot{Plaquette terms are presumably
generated perturbatively but are suppressed by powers of the cutoff.}.

The 2-D sine-Gordon model has the same properties with
respect to the coupling constant $\beta$: it is perturbatively
finite for all $\beta$ but its free-field perturbation theory is
unstable, without additional coupling constant renormalizations, for
$\beta^2 > 8\pi$.
The sine-Gordon field theory is equivalent to the grand
canonical sum of a Coulomb plasma, and the sine-Gordon phase transition
has a nice physical interpretation in terms of the Coulomb gas picture\samuel.
As $\beta$ increases, the free ions of the Coulomb gas, represented by
vertex operators $\exp (\pm \beta \phi )$ in the sine-Gordon model,
collapse to form dipoles and a new gas of interacting dipoles is formed.
This can be interpreted in the sine-Gordon model as the
appearance of a new dimension 2 renormalizable operator
at this fixed point.  One can consider the sine-Gordon model for values of
$\beta^2 > 8\pi$ as long as the additional renormalization for the
new operator is taken into account\agg.

In TLQED, `free ions' are given by fermion charges separated
by one link and connected by a flux tube: $\psi^\dagger \exp (gA ) \psi$.
The `Coulomb gas' in TLQED is then a gas of $e^+$ $e^-$ pairs,
where the charges, separated by a single link, interact
via Coulomb interactions.  The `dipoles' of the strong
coupling phase are pairs of $e^+$ $e^-$ flux tubes,
with strongly interacting photon fields.

The conjecture is that the non-perturbative $g^2_2 (a) = 4\pi$ critical point
of TLQED, where this phase transition occurs, is the critical point of
spontaneous
chiral symmetry breaking, where the $\psibar \psi$ order parameter
gets a vacuum expectation value.
The bare coupling constant $g_2 (a)$ is the `quenched' coupling constant
of TLQED, because fermion loop corrections are obviously not included in the
bare Hamiltonian.  Both the quenched lattice simulations and
the quenched planar approximation exhibit chiral symmetry breaking
for $\alpha_{\rm bare} \sim 1$.
The phase transition in the quenched planar approximation has been
previously compared to the phase transition of the sine-Gordon
model by Miransky\miransky, who interpreted the phase transition
of each model as a collapse phenomenon.  At the critical point
in the quenched planar approximation, the anomalous scaling dimension
of the fermion is 1, and the four-fermion
term becomes a renormalizable operator\llb.
It is
tempting to associate the `dipoles' of the strong coupling phase
of TLQED with renormalizable four-fermion operator of the quenched planar
approximation.
We used the strong coupling expansion of TLQED in section 6 to calculate
explicitly the spontaneous chiral symmetry breaking in the infinite
coupling limit.

Recent lattice gauge theory simulations indicate
that the UV fixed point of chiral symmetry breaking in the quenched theory
may be trivial in the full unquenched theory\dgroup\ddgroup.

Some of the results of this paper can immediately be applied to
more realistic transverse lattice models.  In particular, the construction
of staggered fermions and the analysis of chiral symmetry breaking via
the strong coupling expansion can be easily
generalized to non-abelian gauge theories.

We now briefly mention two formal areas of the theory
that would be interesting to pursue.  In section 4, we noted
that TLQED can be covariantly (in the 2-D sense) bosonized.
Bosonization plays a central role in explaining why the
Schwinger model is exactly soluble.  It would be interesting to understand
the continuum limit of this bosonized version of TLQED.  It
would also be interesting, to work in the `opposite' direction --
to covariantly bosonize transverse
lattice fermions, and then put the two continuous coordinates on a lattice.
Naively, this would generate a 4-D lattice theory where the
fermions are interpreted as bosonic solitons, and the
functional integral over fermions is `gaussian' and easier to simulate\foot{
An idea pointed out to me by W.~Bardeen.}.
Secondly,
TLQED is an interacting 2-D field theory in the form of
a combined Schwinger and sine-Gordon model.  It might be possible
to solve the sine-Gordon `part' by using inverse scattering/Bethe
ansatz methods.  Then the non-integrable Schwinger terms would have to
be treated as perturbations in the space of Bethe-ansatz states.

\ack
I am indebted to W.~Bardeen for clarifying and stimulating discussions.
%\vfill\eject
\listrefs
\bye